\documentclass{iaus}
\usepackage{graphics}

  \checkfont{eurm10}
  \iffontfound
    \IfFileExists{upmath.sty}
      {\typeout{^^JFound AMS Euler Roman fonts on the system,
                   using the 'upmath' package.^^J}%
       \usepackage{upmath}}
      {\typeout{^^JFound AMS Euler Roman fonts on the system, but you
                   dont seem to have the}%
       \typeout{'upmath' package installed. iaus.cls can take advantage
                 of these fonts,^^Jif you use 'upmath' package.^^J}%
      }
  \else
  \fi

  \checkfont{msam10}
  \iffontfound
    \IfFileExists{amssymb.sty}
      {\typeout{^^JFound AMS Symbol fonts on the system, using the
                'amssymb' package.^^J}%
       \usepackage{amssymb}%

      }{}
  \fi

  \IfFileExists{amsbsy.sty}
    {\typeout{^^JFound the 'amsbsy' package on the system, using it.^^J}%
     \usepackage{amsbsy}}
    {\providecommand\boldsymbol[1]{\mbox{\boldmath $##1$}}}

\newsavebox{\astrutbox}
\sbox{\astrutbox}{\rule[-5pt]{0pt}{20pt}}

\newcommand\aj{AJ}
\newcommand\apj{ApJ}
\newcommand\apjs{ApJS}
\newcommand\aap{A\&A}
\newcommand\mnras{MNRAS}

\title[The Interplay among Black Holes, Stars and ISM in Galactic 
       Nuclei]{Why does Low-Luminosity AGN Fueling Remain an Unsolved Problem?}

\author[Paul Martini]%
{Paul Martini}%

\affiliation{Harvard-Smithsonian Center for Astrophysics; 60 Garden Street, 
MS20; \\ Cambridge, MA 02138, USA; pmartini@cfa.harvard.edu}

\pubyear{2004}
\volume{222}
\pagerange{1--7}
\date{April 9, 2004 and in revised form ??}
\jname{The Interplay among Black Holes, Stars and ISM \\in Galactic Nuclei}
\editors{Th. Storchi Bergmann, L.C. Ho \& H.R. Schmitt, eds.}
\begin{document}

\maketitle

\begin{abstract}
Despite many years of effort, observational studies have not found a strong
correlation between the presence of any proposed fueling mechanism and
low-luminosity AGN.
After a discussion of the mass requirements for fueling, I summarize this
observational work and provide a number of hypotheses for why the nature of
AGN fueling has remained unresolved. In particular, I stress the potential
importance of the increasing number of candidate fueling mechanisms with
decreasing mass accretion rate, the relevant spatial scales for different
fueling mechanisms, and the lifetime of an individual episode of
nuclear accretion. The episodic AGN lifetime is a particularly relevant
complication if it is comparable to or shorter than the time that the
responsible fueling mechanisms are observationally detectable. I conclude
with a number of relatively accessible areas for future investigation.
\end{abstract}

\firstsection 

\section{Introduction}

One of the main, unsolved problems in AGN research is how the AGN fuel is 
transported to the central, supermassive black hole. For low-luminosity 
AGN, the likely source of this fuel is the host galaxy itself, and in 
particular the ISM. As most of this material is distributed in a rotating 
disk extending out to kiloparsec scales, the problem of AGN fueling is 
essentially a problem of angular momentum transport: All but approximately 
one part in $10^7$ of the angular momentum must be removed for material 
to flow from kiloparsec scales to the event horizon on AU scales. 

The problem has remained unsolved in part because the most relevant spatial 
scales for fueling, the central parsec and inward, are only observable in 
the very nearest galaxies. 
An angular resolution of $0.1''$ only corresponds to 1~pc or better spatial 
resolution out to a distance of 2~Mpc. 
Even the central 10~pc are not readily resolved for large samples of nearby 
AGN. 
The present technical limitations on angular resolution have therefore made it 
impossible to study AGN fueling directly and instead driven investigations 
to indirect, statistical studies of active and inactive galaxy samples in 
order to identify the mechanism(s) responsible for fueling accretion. 

\subsection{Requirements for Fueling} 

Before consideration of these investigations, it is valuable to first 
consider the mass accretion rates estimated to produce the observed 
population of low-luminosity AGN, which are defined here to be any AGN 
less luminous than a QSO. 
The luminosity of an AGN is related to the mass accretion rate by 
$L = \epsilon \dot{M} c^2$, where $\epsilon$ is the radiative efficiency 
of the accretion. This accretion is commonly assumed to occur via a 
Shakura-Sunyaev thin disk (Shakura \& Sunyaev 1973) with a constant radiative 
efficiency $\epsilon = 0.1$. 
To maintain Eddington accretion onto a supermassive black hole requires a 
mass accretion rate of
\begin{equation}
\dot{M}_{Edd} = 0.2 \epsilon^{-1}_{0.1} \left( \frac{M}{10^7 {\rm M}_{\odot}} \right) {\rm M}_{\odot} {\rm yr}^{-1} 
\end{equation}
where $\epsilon_{0.1} = \epsilon/0.1$. The mass accretion rate is commonly 
parametrized in terms of the Eddington rate as the dimensionless accretion 
rate $\dot{m} \equiv \dot{M}/\dot{M}_{Edd}$, similarly the bolometric 
luminosity can be expressed as $l \equiv L_{bol}/L_{Edd}$. For a standard 
accretion disk $l = \dot{m}$, while lower efficiency ADAF models,  
which likely become important below $\dot{m} = \dot{m}_{crit} \sim 0.01$, 
predict $l \propto \dot{m}^2$ from a scaling of 
$\epsilon_{0.1} = \dot{m}/\dot{m}_{crit}$ 
(Narayan, Mahadevan, \& Quataert 1998). 
The required mass accretion rates from fueling may be even higher if 
ADIOS/CDAF models are important, as these models predict that only a small 
fraction of the supplied mass is actually accreted by the black hole 
(Blandford \& Begelman 1999). 
The key importance of these relations is that the mass accretion rates 
required for low-luminosity AGN do not decline as rapidly as their luminosity. 

The most luminous low-luminosity AGN, Seyfert 1s, appear to have 
central, supermassive black holes with $M_\bullet = 10^7 {\rm M}_{\odot}$ 
(Ferrarese et al.\ 2001) and 
$l = 0.1$, which corresponds to accretion at $\dot{m} = 0.1$ or 
$\dot{M} = 0.02\,{\rm M}_{\odot}$/yr. 
More typical Seyferts, which constitute a total 
of approximately 10\% of the luminous galaxy population (Ho, Filippenko, 
\& Sargent 1997)
likely accrete with $\dot{m} = 0.01$, while the 30\% of the luminous galaxy 
population that are LINERs likely accrete with $\dot{m} = 10^{-2} - 10^{-4}$ 
based on estimates of $l$ (Ho 1999). 
If all luminous galaxies go through periodic episodes as AGN, then over a time 
period of $10^8$ years\footnote{This time period is only adopted to illustrate 
the potential total fuel requirements of a low-luminosity AGN. The relative 
numbers of Seyferts and LINERs only constrain the duty cycle of these phases 
and not the lifetime}, a $10^7 {\rm M}_{\odot}$ black hole will appear as a 
Seyfert galaxy for a total of $10^7$ yr and accrete on order $10^{3} 
{\rm M}_{\odot}$, appear to be a LINER for $3 \times 10^7$ yr and accrete 
$10^{3} {\rm M}_{\odot}$, and appear to be inactive for $6 \times 10^7$ yr. 
Therefore the mass inflow rates and total mass reservoirs required to power 
low-luminosity AGN are relatively meager. 

\subsection{Proposed Fueling Mechanisms} 

Figure~1 lists the many mechanisms proposed to drive angular momentum 
transport in the host galaxy and provide fuel to the central parsec. 
These mechanisms can be divided between gravitational and hydrodynamic 
mechanisms. Gravitational mechanisms, such as galaxy interactions 
(Toomre \& Toomre 1972) and large-scale bars (Simkin, Su, \& Schwarz 1980), 
remove angular 
momentum through torques, while hydrodynamic mechanisms, such as turbulence 
in the ISM (Elmegreen et al.\ 1998), remove angular momentum through gas 
dynamical effects. 
Many of these mechanisms are discussed in the review by Shlosman, Begelman, 
\& Frank (1990) and the more recent review by Wada (2004). 
The latter in particular provides 
an excellent overview of recent theoretical work on hydrodynamic fueling 
mechanisms and places particular emphasis on high resolution simulations of 
the multiphase ISM in the central kiloparsec. 

\begin{figure}
\includegraphics{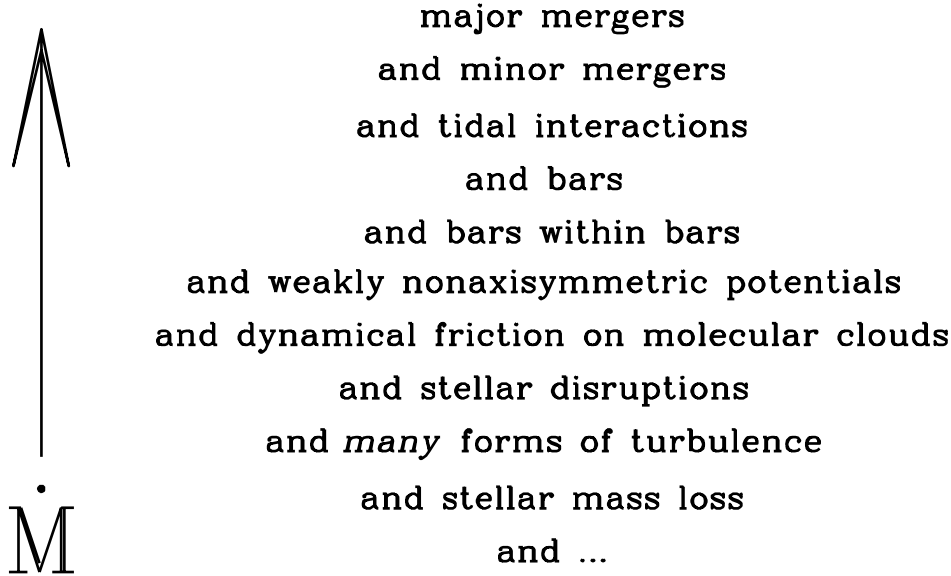}
\caption{\small
Mechanisms proposed to fuel accretion onto black holes at the centers of 
galaxies. The fueling mechanisms are approximately ordered by 
their expected maximum accretion rate. Progressively larger numbers of 
progressively more common mechanisms may be responsible for 
supplying the lowest rates of accretion. This suggests that while the relevant 
question for luminous AGN may be ``Why are active galaxies active?'' the 
question should be ``Why are inactive galaxies inactive?'' at the 
lowest accretion rates. 
}
\end{figure}

While a large number of mechanisms have been proposed for AGN fueling, the 
vast majority are only likely to provide relatively low mass accretion rates. 
Mergers between galaxies, particularly major mergers, is the mechanism most 
commonly invoked to explain the high accretion rates required to power luminous 
QSOs, while at the somewhat lower mass accretion rates responsible for 
Seyfert-level luminosities, mechanisms such as bars and minor mergers have 
also been considered. A progressively larger number of candidate mechanisms 
could be important at yet lower mass accretion rates. The mechanisms listed 
in Figure~1 are approximately ordered by their relative {\it maximum} mass 
accretion rates.  As the mass accretion rate to the nucleus is expected to
gradually taper off, rather than stop abruptly, mechanisms invoked to explain
high mass accretion rates may also produce low mass accretion rates as the 
fuel supply is gradually depleted. 

\subsection{Observational Searches} 

As the spatial scales most important for fueling are currently inaccessible 
in most galaxies, searches have instead employed large samples of AGN and 
inactive, control galaxies to identify the fueling mechanism(s). 
Ideally, a candidate fueling mechanism should only be found in the AGN sample. 
These experiments are challenging in that they require both large samples 
of galaxies and very careful control of systematic effects in the selection 
of both the AGN and the control samples. Claims over the years of 
statistically significant excesses have all diminished after reevaluation 
of sample sizes and the details of the AGN and control sample 
selection (e.g.\ different distance distributions, host galaxy types). 
Sample selection remains a very important consideration as progressively 
more AGN are found in progressively more sensitive surveys (e.g.\ Ho et al.\ 
1997) 
and few studies of AGN fueling use samples selected with uniform and 
unbiased criteria, such as hard X-ray luminosity, nor do they have comparably 
sensitive observations of the inactive, control sample. 

The two main and most readily observed `large-scale' mechanisms are 
bars and interactions. Neither of these features is seen in significant excess 
in AGN samples compared to carefully-matched control samples 
(Fuentes-Williams \& Stocke 1988; Mulchaey \& Regan 1997; Schmitt 2001; 
see also Schmitt, {\it these proceedings}). 
One explanation for these results is that AGN fueling is predominantly 
mitigated by smaller-scale phenomenon than accessible in these ground-based 
surveys (e.g.\ Martini \& Pogge 1999), a point illustrated in part by Figure~1. 
While the fueling mechanisms 
are approximately ordered by accretion rate, the mechanisms proposed to 
produce lower accretion rates are also progressively smaller-scale phenomena 
and more difficult to identify observationally. The desire for finer 
spatial resolution motivated a careful study of the circumnuclear region 
(on 100 pc scales) with {\it HST}, but still no significant differences 
between AGN and control samples were found (Martini et al.\ 2003a). 

\section{Why have surveys been unsuccessful?} 

The null results from searches for AGN fueling mechanism(s) have successfully 
eliminated many simple models for AGN fueling, such as the importance of a 
single mechanism and the hypothesis that the AGN lifetime is comparable to the 
Hubble time. Several relatively straightforward scenarios that can explain 
the present null results include the importance of multiple fueling mechanisms, 
the overly broad classifications for fueling mechanisms, correlations between 
fueling mechanism and fueling rate, and the importance of time dependence. Here 
I discuss each of these scenarios in turn and describe both how they can 
explain current observations and be tested by future work. 

{\bf 1. Multiple fueling mechanisms are important:} This scenario is the 
simplest explanation of current observational results. If more than one 
mechanism is important (e.g.\ bars and interactions), then the significance 
of the correlation between any one mechanism and AGN will be 
diluted by AGN fueled by other mechanisms. There is good reason to consider 
this possibility, as the theoretical motivations for multiple mechanisms 
are equally plausible. This scenario could be tested with the 
`classical' survey approach. 

{\bf 2. Current classifications for fueling mechanisms are too broad:} 
Observational work has already shown that there is a wide dispersion in the 
detailed properties of mechanisms proposed for fueling, such as the fact that 
not all barred galaxies possess offset shocks that connect to grand-design 
nuclear spiral structure (Martini et al.\ 2003b). If only certain subclasses of 
a candidate fueling mechanism actually remove angular momentum (e.g.\ only 
certain types of bars), then the correlation between this broader class and 
AGN will be diluted. This rationale could also explain the absence of AGN in 
some barred and/or interacting galaxies. Reevaluation of surveys for the 
incidence of certain types of bars, for example only strong bars, is a 
simple test of this scenario. 

{\bf 3. There are correlations between fueling mechanism and fueling rate:} 
As noted above in the discussion of Figure~1, the number of proposed fueling 
mechanisms 
increases as the required accretion rates diminish. Present surveys do not 
explicitly consider low-luminosity AGN with a specific range of $\dot{M}$ 
or $\dot{m}$. For example, Figure~1 suggests that it might be easier to 
identify the fueling mechanism for higher accretion rates, both because the 
relevant mechanisms may be fewer in number and they are visible on larger 
scales. 
Thus inclusion of all low-luminosity AGN would dilute any potential signal for 
the high accretion rate subsample. The importance of correlations between 
fueling mechanisms and fueling rates could be tested by specifically 
identifying samples of higher accretion rate AGN with empirical techniques 
to estimate $M_{\bullet}$ combined with multiwavelength observations to 
estimate $L_{bol}$. The main difficulty of this approach is that more luminous, 
higher accretion rate objects are rarer and on average further away. The 
observed spatial resolution is thus poorer, in addition to the greater 
contamination of the circumnuclear region by the brighter nuclear point source. 

Lower accretion rate objects are progressively more challenging to study 
due to the larger pool of candidate mechanisms, the difficulty in 
quantitatively classifying many of the proposed fueling mechanisms
(e.g.\ nonaxisymmetric potentials, dynamical friction on molecular clouds), 
and the small spatial scales relevant for most of the mechanisms. 
Lower mass accretion rates in particular are likely to be controlled by 
small-scale, essentially stochastic processes such as the distribution of 
molecular clouds and the time since the last supernova. 
For these largely hydrodynamic processes, Wada (2004) finds the accretion 
rate into the nuclear region ($<1$pc) varies over three orders of magnitude on 
a fluctuation timescale of $10^4 - 10^5$yr due to the inhomogeneity of the 
circumnuclear region. At the lowest mass accretion rates, no correlation 
between various fueling mechanisms and nuclear activity may be found unless 
the central parsec can be resolved. 

{\bf 4. Time dependence is important:} The above three methods illustrate minor 
complications that can plausibly explain why existing surveys to identify the 
AGN fueling mechanism have been unsuccessful. These complications can in 
principle be tested with more careful sample selection and classification of 
the various mechanisms proposed for AGN fueling. However, an alternate, 
physically plausible explanation for the null results to date is that time 
dependence is important. If the AGN lifetime is comparable to or shorter than 
the inflow time due to a given fueling mechanism, studies of the frequency of 
some fueling mechanism(s) in active and inactive galaxies will show no 
correlation with activity, even if the correct fueling mechanism has been 
identified.  

\begin{figure}
\includegraphics{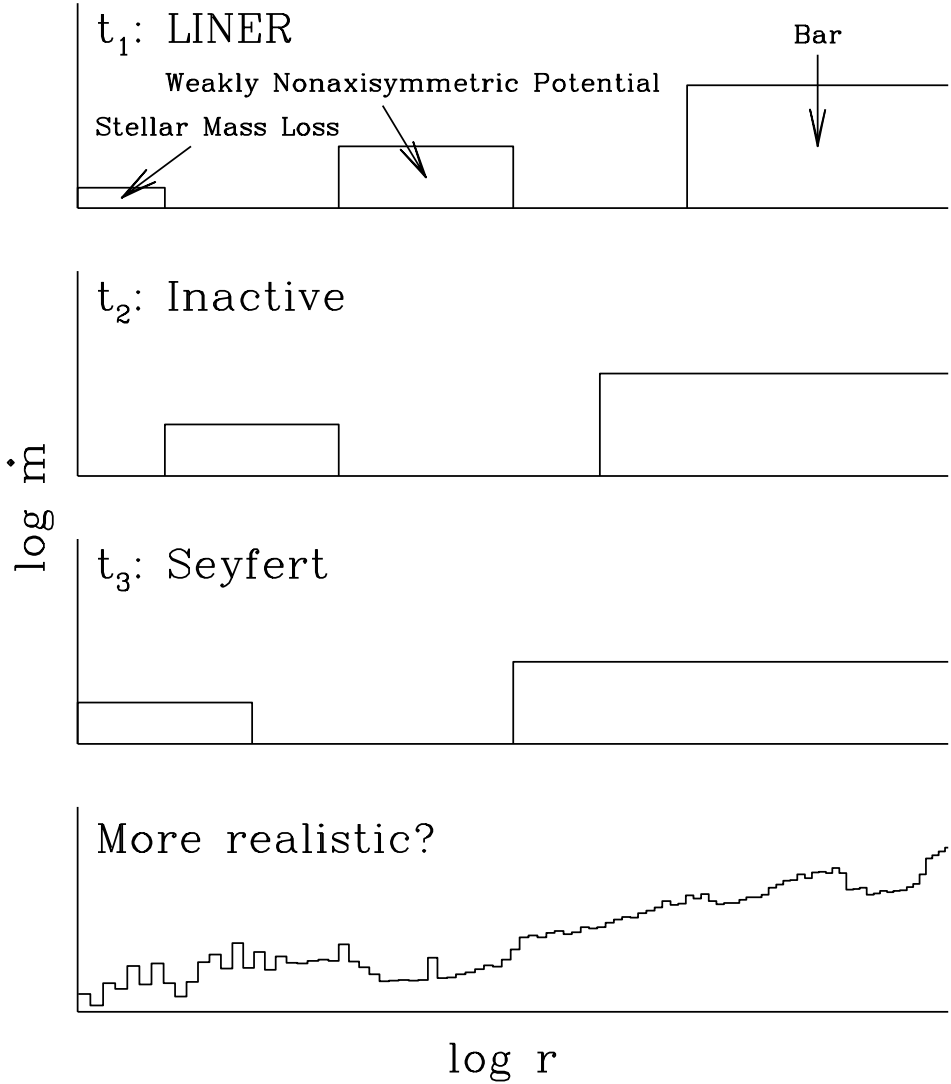} 
\caption{\small
Simple cartoon illustrating the potential importance of time evolution in AGN 
fueling. The top three panels present snapshots of the mass accretion 
rate as a function of radius at three separate times and illustrate how 
the presence of fueling mechanisms on larger spatial scales may be 
decoupled from the presence of an AGN if the episodic AGN timescale is 
comparable to or shorter than the inflow timescale due to the relevant fueling 
mechanism. The bottom panel illustrates the likely complexity of inflow rate 
as a function of radius for an actual galaxy. 
}
\end{figure}

The potential importance of time-dependence is illustrated in Figure~2 with 
a simple cartoon of azimuthally-averaged inflow rate as a function of 
radius. Here I have assumed that several different mechanisms are driving 
inflow at a range of different galactic radii. At time $t_1$, 
mass loss from stars provides fuel to produce the spectral signatures 
of a LINER, while at larger scales higher inflow rates are produced by 
gravitational torques. By some later time $t_2$, accretion in the nuclear 
region has switched off and no AGN is visible, yet the gravitational torques 
continue to drive inflow toward the galactic nucleus. At this time, the galaxy 
would be classified as barred, yet not as an AGN. Eventually, at time $t_3$ 
the greater level of inflow from the gravitational torques reaches the 
central black hole and produces a sufficiently luminous AGN for the galaxy 
to be classified as a Seyfert. 
Without providing actual units for time, mass accretion rate, 
or radius, the progression from $t_1 - t_3$ in this example illustrates how 
the classical approach to the identification of the AGN fueling mechanism(s) 
could be completely thwarted if the episodic lifetime of an AGN is comparable 
to the lifetime of one or more important fueling mechanisms. The final 
panel is intended to illustrate the likely complexity of the actual 
phenomenon, rather than an actual physical scenario. 

\section{Future Directions}

The observational studies discussed here have eliminated many simple models for 
AGN fueling and led to a much clearer appreciation of the complexity of 
the problem. Current and future observational facilities, new empirical 
methods to estimate black hole masses, as well as computational improvements, 
enable several new lines of investigation. 

\indent
{\bf Detailed studies of the circumnuclear region:} The observational 
studies described above have revealed significant correlations between the 
circumnuclear dust morphology and the presence or absence of a large-scale bar. 
Further work to understand the formation of nuclear spiral structure would 
be of great value, particularly measurements of the pitch angle 
and mass surface density with radius, as well as the distribution of 
molecular clouds. Kinematic data for many disk galaxies with higher spatial 
resolution is also needed, both to measure rotation curves and hence the mass 
distribution at small physical scales for accurate theoretical models, and 
also to quantify the extent to which the gas kinematics in the centers of 
galaxies are chaotic. While the best resolution will be obtained with 
future ALMA observations, significant progress will be possible in the near 
future with CARMA and the SMA, as well as with the completion of current 
surveys of nearby galaxies such as NUGA (Combes et al.\ 2004). 

\indent
{\bf Physically defined samples:}  The existence of empirical techniques to 
estimate black hole masses make it possible to reconsider AGN fueling with 
samples defined by $M_{\bullet}$ and $L_{bol}/L_{Edd}$. In particular, 
success in identifying strong evidence for any AGN fueling mechanism may first 
come from investigation of samples with higher mass accretion rates. This 
work, and all surveys that employ the classical approach of AGN and inactive, 
control samples, also require extremely careful selection of the samples. 
In particular, it is extremely important to have an unbiased and uniform 
selection of both the AGN and control sample, for example based on 
hard X-ray luminosity. 

\indent
{\bf Improved hydrodynamic simulations of the circumnuclear regions:} 
Current observations can probe AGN with a spatial resolution as fine as tens of 
parsecs. This spatial resolution is comparable or only slightly worse than 
that achieved by current multiphase ISM models on galaxy scales (e.g. Wada, 
{\it these proceedings}). 
Increased spatial resolution in the simulations, along with estimates of 
the mass accretion rates at different spatial scales and any correlations 
with observable fueling mechanisms, would be very useful for interpreting 
the present observations. 
Reconsideration of mass inflow in bars and mergers in models with high 
spatial resolution and self-consistent star formation would also be of 
great value, as would velocity field calculations for the interpretation of 
two-dimensional kinematic observations, and further consideration of the time 
domain. 

\vspace{0.05in}

While surveys to identify the mechanisms responsible for AGN fueling have 
provided a string of null results, the lessons learned from these 
investigations provide several clear and observationally testable hypotheses.  
New work based on large samples with empirical estimates of black hole masses, 
multiwavelength measurements of $L_{bol}$, unbiased selection criteria, and 
careful definition of candidate fueling mechanisms offers the promise of 
significant progress toward solving the problem of how low-luminosity AGN 
are fueled. 

\begin{acknowledgments}
I acknowledge support from HST grant AR-9547, the International Astronomical 
Union, and the conference organizers. I would also like to thank Rick Pogge 
for helpful comments and discussions. 
\end{acknowledgments}

\end{document}